\documentstyle[psfig,epsf,conf-X]{article}
\begin{document} 
\small
\heading{%
%Begin Heading
%
The hard X--ray spectrum of Compton--thick Seyfert 2 galaxies and the
synthesis of the XRB
% End Heading
}
\par\medskip\noindent
\author{%
%Begin Author names
Giorgio Matt$^1$, Fulvio Pompilio$^1$,  Fabio La Franca$^1$
%End Author names
}
\address{%
%First address
Dipartimento di Fisica, Universit\`a Roma Tre, Via della Vasca Navale 84,
I--00146 Roma, Italy.
}

\begin{abstract}
A synthesis model for the cosmic X--ray
Background (XRB) is presented, 
which includes a proper treatment of Compton scattering 
in the absorbing matter for type 2 AGN. Evidence for a decrease  of the 
relative importance of type 2 AGN at high redshift is found, which may be due
either to a decrease of the relative number of obscured sources, or (more
plausibly) to an increase of the fraction of Compton--thick absorbed sources. 
The XRB spectrum, soft X--rays and hard X--rays source counts can be
simultaneously fitted only if the XRB normalization as derived 
from BeppoSAX/MECS measurements (\cite{v99}) is adopted.
\end{abstract}

\section{Modeling the X--ray absorption in Compton--thick source }

BeppoSAX observations have shown that a large fraction (at least 
50\%) of nearby Seyfert 2 galaxies are Compton--thick, i.e. the nucleus
is obscured by matter with $N_H \geq \sigma_T^{-1} = 
1.5\times10^{24}$ cm$^{-2}$ (\cite{m98}, \cite{rms}). 
Because Seyfert 2s outnumber Seyfert 1s by a large factor, this means that 
Compton--thick Seyfert 2s are the most common type of AGN 
in the local Universe. It is possible that heavily obscured
sources were even more common in the past (\cite{f99}), and then
Compton--thick sources should be an important ingredient in XRB
synthesis models, despite their low flux. 

It is therefore important to model in detail the X--ray spectrum emerging
from such a thick absorber. We have calculated transmitted spectra
by means of Monte Carlo simulations assuming
a spherical geometry, with the X--ray source in the centre.
All relevant physical processes: photoelectric absorption, Compton scattering 
(with fully relativistic
treatment),  and fluorescence (for iron atoms only), have been
included in the code. More details can be found in \cite{mpl}. 

To illustrate the importance of a proper treatment of the transmission 
spectrum, in Figure 1 the case for $N_H=3\times10^{24}$ cm$^{-2}$ is shown.
For comparison, we also plot the spectrum obtained with only
photoelectric absorption (an unphysical 
situation) and photoelectric plus Compton absorption (neglecting
scattering, which corresponds to obscuring matter with small covering
factor, an unlikely situation given the large fraction of Compton--thick
sources). The differences between the three curves are large, and
fitting real data with the inappropriate model can make a big
difference in the derived parameters. Let us discuss as an example the
case of the Circinus Galaxy. The BeppoSAX observation 
revealed the nuclear radiation transmitted through a Compton--thick 
absorber (\cite{m99}). When fitted with the ``small cloud" absorber,
the best fit value for
the column density is 6.9$\times$10$^{24}$ cm$^{-2}$, and the 2--10
keV extrapolated luminosity is 1.5$\times$10$^{44}$ erg s$^{-1}$, a surprisingly
large value when compared to the IR luminosity. If the spectrum is instead
fitted with the spherical model (then including Compton scattering), the
column density is 4.3$\times$10$^{24}$ cm$^{-2}$, and the 2--10
keV luminosity reduces to a much more reasonable value of  10$^{42}$ 
erg s$^{-1}$.

\begin{figure}
\mbox{\epsfxsize=7cm \epsffile{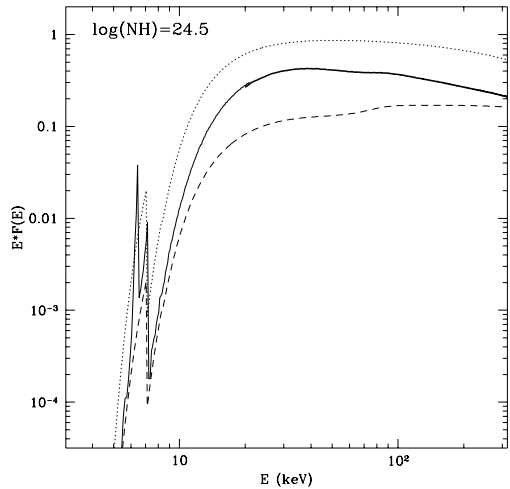}}
\caption[]{The X--ray transmitted spectrum for 
$N_H=3\times10^{24}$ cm$^{-2}$. The solid line is for the complete spherical
model discussed in \cite{mpl}. The dashed line refers to absorption by 
matter with small covering factor, while the dotted line is computed in the
unphysical assumption of only photoelectric absortion, and is shown here
only for the sake of illustration.}
\end{figure}

\section{ The XRB synthesis model and the evolution of AGN}

We developed a synthesis model for the XRB 
based on the standard assumption that
the XRB is mostly made by a combination of type 1 and 2 AGN (\cite{sw},
\cite{com} and references therein). 
Below we schematically summarize the main assumptions of the model.
Further details can be found in \cite{plm}.

\begin{enumerate}
\item AGN spectra
\begin{enumerate}
\item type 1 (AGN1) spectrum:
\begin{itemize}
\item power law ($\alpha =0.9$) + exponential cut-off (${E}_{c}=400$ keV);
\item Compton reflection component (accretion disk, ${\theta }_{obs}\sim
{60}^{\circ }$);
\end{itemize}
\item type 2 (AGN2) spectrum (\cite{mpl}):
\begin{itemize}
\item primary AGN1 spectrum obscured by cold matter:\par ${10}^{21}\le
{N}_{H}\le {10}^{25}{cm}^{-2}$, $\frac{dN(log{N}_{H})}{d(log{N}_{H})}\propto
log{N}_{H}$;
\item Compton scattering within the absorbing matter fully included.
\end{itemize}
\end{enumerate}

\item Cosmological parameters 
\begin{enumerate}
\item PLE (${\Phi }^{*}(z=0)=1.45\times {10}^{-6}
{Mpc}^{-3}{({10}^{44}erg~{s}^{-1})}^{-1}$);
\item power law evolution for the break-luminosity:\par
 ${L}^{*}(z)\propto
{(1+z)}^{k}$ up to ${z}_{max}=1.73$, with \par
${L}^{*}(z=0)=3.9\times {10}^{43}
erg~{s}^{-1}$ and $k=2.9$ (model H of \cite{b94});
\item the redshift integration is performed up to ${z}_{d}=4.5$.
\end{enumerate}
\end{enumerate}

\subsection{The $R(z)$ model}

The best-fit to the high energy (3-50 keV) XRB HEAO-1 data (\cite{m80}),
but with a $\sim $30 \% higher normalization (according to the
BeppoSAX/MECS results below 10 keV, \cite{v99})
is performed by a ${\chi}^{2}$-minimization procedure.
The inclusion of a redshift--dependent term in the number ratio of
the two types of sources (i.e. $R(z)= N(type2,z)/N(type1,z)$)
results in an improvement of the fit
at the 99\% confidence level.
Our best solution is shown in Figure 2 and is described
by the following analytical form:
\[R(z)={R}_{0}\times {(1+z)}^{{k}_{1}}{e}^{{k}_{2}z}\]
with ${R}_{0}=4$ (according to \cite{m98}). The best-fit
parameters are ${k}_{1}=2.8\pm 0.2$ and ${k}_{2}=-1.5\pm 0.1$. 
It is worth noticing that this result does not necessarily imply that the 
actual number of Seyfert 2s diminishes with $z$, 
but rather that their contribution
to the XRB diminishes. This may well be due to a larger fraction of obscured
objects being Compton--thick in the past, as proposed by \cite{f99}. 
Hopefully, {\it Chandra} and XMM surveys will be able to check this 
intriguing possibility. 

\begin{figure}[t]
\mbox{\epsfxsize=8cm \epsffile{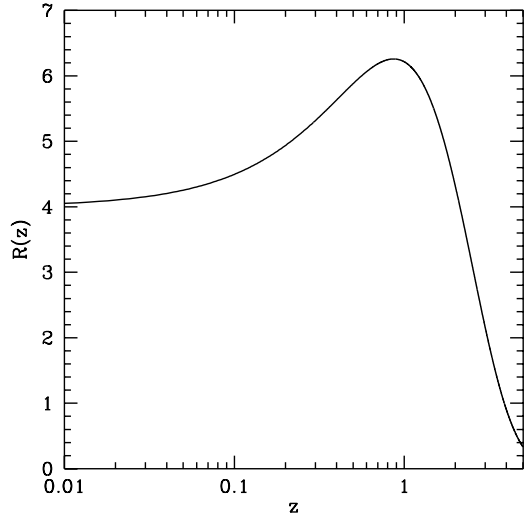}}
\caption[]{The type 2/type 1 number ratio as a function of redshift.}
\end{figure}

\subsection{The source counts}

While there is no much difference in the goodness of the fit 
to both the XRB spectrum and the soft X--ray source counts if the HEAO--1 
normalization is or not readjusted to match the BeppoSAX results, 
it makes a large difference in the
fitting of the hard X--ray source counts. In particular, the 
BeppoSAX (\cite{fiore} and  Comastri, this conference) 5--10 keV source 
counts can be simultaneously 
fitted by our model only if the higher normalization is used. 

\acknowledgements{We thank the HELLAS group for many useful 
and stimulating discussions. We acknowledge financial 
support from ASI and from MURST (grant {\sc cofin}98--02--32.)}

\end{document}